\documentclass[useAMS,usenatbib]{./mn2e}
\usepackage{./psfig}
\input ./journals.def
\input ./abbreviations.def

\title[Intermediate Mass Black Hole X-ray Binaries]{ Intermediate Mass
Black Holes in Accreting Binaries: Formation, Evolution and
Observational Appearance}

\author[Portegies Zwart et al]
       {
	Simon F.\ Portegies Zwart$^{1, 2}$
	Jasinta Dewi$^{3}$
	and
	Tom Maccarone$^{1}$\\
$^1$ Astronomical Institute 'Anton Pannekoek', 
		Univeristy of Amsterdam, Kruislaan 403 \\
$^2$ Section Computational Science,
		 Univeristy of Amsterdam, Kruislaan 403 \\
$^3$ Department of Astrophysics, University of Oxford, 
                 Keble Road, Oxford OX1 3RH, UK \\
}
\begin{document}

\date{Accepted 2145 December 32. Received 1609 December -1; in
original form 1900 October 11}

\maketitle
\pagerange{\pageref{firstpage}--\pageref{lastpage}} \pubyear{2004}

\maketitle

\label{firstpage}

\begin{abstract}
We study the origin of the ultraluminous X-ray source M82-X1 in the
nearby starburst galaxy M82. This X-ray source is of particular
interest as it is currently the best candidate intermediate mass black
hole; it is associated with a 54\,mHz quasi periodic oscillations
with a relatively low ($\sim 1$keV) black-body temperature.
We perform detailed binary evolution calculations of 2--15\,\msun\,
stars which transfer mass to a 100--2000\,\msun\, black hole and
present an empirical model for the X-ray characteristics expected for
such binaries. Based on the binary evolution calculations and the
assumption in our simulations we conclude that the most likely
candidate for the bright X-ray source M82-X1 is a 10-15\,\msun\, star
near the end of its main-sequence or slightly evolved, which transfers
mass to a $\sim 1000$\,\msun\, black hole. We expect the system to be
in the high/soft state.  In that case the binary will not be visible
as a source of gravitational wave radiation, but other transient X-ray
binaries with lower mass donors way be rather bright sources of
gravitational wave radiation.
\end{abstract}


\section{Introduction}

Black holes in the universe have either masses less than about 20
M$_\odot$ (Orosz 2003), \nocite{2003IAUS..212..365O} or masses
exceeding about $10^6$\,\msun\, (see Kormendy \& Richstone, 1995, for
some exceptions see Fillipenko \& Ho 2003; Greene \& Ho 2004). The
first category of black holes are typically the member of a binary
system in a normal galactic environment, the other type of
supermassive black holes can be very bright in X, but always live in
the nuclei of galaxies.  The discovery of bright X-ray sources which
are not in the nuclei of galaxies has led to the suggestion that these
sources may hide black holes, with a mass intermediate between
20\,\msun\, and $10^6$\,\msun; we call them intermediate mass black
holes (IMBH).

Bright non-nuclear X-ray sources have been known to exist since the
1980s (Watson, Stanger \& Griffiths 1984),\nocite{1984ApJ...286..144W}
recently with the advent of Chandra and XMM these sources have
attracted considerably more attention.  The conventional definiton of
an ultraluminous X-ray source is $L_X \apgt10^{39}$ ergs/s, but for
the purpose of this paper we consider only luminosities of $L_X
\apgt10^{40}$ ergs/s as true ULX. This definition corresponds to the
Eddington luminosity of an accreting compact object of about 80 times
the Chandrasekhar mass.

The brightest Galactic X-ray binary, GRS 1915+105, occasionally
reaches $7\times10^{39}$ ergs/s, and should then, according to the
earlier definition be classified as a ULX.  GRS~1915+105, however,
contains a rather ordinary $14\pm4 M_\odot$ black hole (Harlaftis \&
Greiner 2004), which has an Eddington limit of 3.5$\times10^{39}$
ergs/s if it accretes helium, and a factor of two lower if it accretes
hydrogen rich material. We therefore argue that the term ultraluminous
is inflationary, and opt for a more stringent definition (see also
Irwin, Bregman \& Athey 2004; Irwin et al. 2004).  The definition with
$10^{40}$\,erg/s is thus chosen because it is difficult to explain
these sources with conventional X-ray binaries containing stellar mass
black holes (see also Kalogera et al.\, 2004).

Just the X-ray luminosity is not sufficient to identify a ULX as an
IMBH.  A few studies have claimed spectral evidence for intermediate
mass black holes, in the form of relatively low temperature accretion
disks (i.e. $k_B$$T<<1$ keV) which cannot produce the observed high
X-ray luminosities unless the accretion radius is quite large, and
hence flowing into massive black holes (Miller et al. 2003a,b).  On
the other hand, such spectral results are generally based on fitting
the disk emission (Shakura \& Sunayev 1973) using the models of
Mistuda et al. (1984) and Gierlinski et al. (1999) including a power
law, which presumably fits the Compton up-scattered emission at high
energies (Thorne \& Price 1975; Shapiro, Lightman \& Eardley 1976;
Sunyaev \& Tr\"umper 1979).  Inferring the black hole mass from the
inner disk radius measured by spectral fits typically works well for
sources whose X-ray fluxes are dominated by a quasi-thermal component,
but less well for sources dominated by the power law component (see
e.g. Sobczak et al. 2000).  As the ULXs for the spectral mass
measurement technique has been applied have spectra dominated by the
power law component, these results should be taken with some caution.

A small subset of extragalactic ULXs show properties that are unlike
X-ray sources in the Milky Way.  These objects provide some indication
for the existence of either intermediate mass black holes or for
unusual modes of accretion.  Many of these bright X-ray sources are
variable, which helps provide confidence that the emission comes from
a single accreting object, and some also show evidence of associations
with counterparts in other wavelengths, such as star clusters (Zezas
\& Fabbiano 2002; Smith \& Wilson 2003)\nocite{2002ApJ...577..726Z}
\nocite{2003ApJ...591..138S} or optical emission nebulae (Shih,
Iwasawa \& Fabian 2003)\nocite{2003MNRAS.341..973S} or even bright
stars (Liu, Bregman, Seitzer 2004), which makes them unlikely to be
background objects.

There are a number of theoretical models for these sources which do
not require an intermediate mass black hole, but rather use the fact
that the Eddington limit applies strictly to spherical accretion.  The
non-IMBH explanations include mechanical beaming (King et al. 2001),
relativistic beaming from a jet (K\"ording, Falcke \& Markoff 2002) or
photon bubble instabilities (Begelman 2002); many of these
configurations can be obtained from a rather ordinary 10\,\msun\,
black hole accreting from a donor during case B or C mass transfer
(Podsiadlowski et al. 2003).  A review of the theory of ULXs can be
found in King (2003).

One particular source, M~82~X-1, represents an especially compelling
candidate for hosting an intermediate mass black hole (Ebisuzaki et al
2001; Portegies Zwart et al.\, 2004).  It is located sufficiently
close to a massive young star cluster that a chance superposition is
unlikely, so it can be claimed to truly reside in M~82 with
considerable confidence.  It has an X-ray luminosity of about
5$\times10^{40}$ ergs/s (i.e. well in excess of the luminosity of any
Galactic X-ray binary), and shows a 54 mHz quasi-periodic oscillation
(QPO) as well as a broad iron line (Strohmayer \& Mushotzky 2003).
They argue that in a mechanical beaming scenario, the QPO should be
broadened by scatterings which increase the photons' light travel
times, while the iron line should be narrow, since lines are typically
broadened due to projected rotation effects.  In a relativistic
beaming scenario, neither QPOs nor iron lines are expected.
Furthermore, they note that the observed QPO and noise spectrum are
reminiscent of the Galactic X-ray binaries in their very high state
(see van der Klis 1995 for a review), except that the QPO frequency is
a factor of about 10-100 lower than those seen in the Galactic X-ray
binaries.  Applying a simple linear mass scaling to the QPO frequency
implies then a mass of about 100-1000 $M_\odot$ for the black hole.

For a characterization of the bright X-ray sources it is important to
identify models for their origin and evolution. For the remainder of
the paper we adopt the intermediate mass black hole hypothesis and
study the characteristics of an X-ray binary in which such a black
hole accretes from a stellar companion. Our assumptions are motivated
by the compelling case of M~82~X-1.  The configuration in mind can be
compared to low-mass X-ray binaries in which a main-sequence star or
giant fills its Roche-lobe to a black hole, in this case, however, the
accreting object is a black hole {\large O}(1000)\,\msun.


In this paper we discuss the consequence of a $\sim 1000\msun$ black
hole that accretes from a stellar companion.  We envision that such
systems may form in the dense cores for young star clusters.  We
discuss the formation of such systems in Sect.\,\ref{Sect:FandE}.

We further evolve the binary system through several phases of
Roche-lobe overflow (RLOF).  For this purpose we perform detailed
binary evolution calculations for such systems and demonstrate that if
the binary can form with sufficiently small orbital period a stable
low-mass X-ray binary forms in which the donor is slowly consumed by
the black hole. The X-ray emission while the $\apgt 5$\,\msun\, donor
is on the main-sequence reaches luminosities above the Eddington limit
for a 1\,\msun\, accreting compact object. In the shorter post
main-sequence phase less massive donors also exhibit super-Eddington
luminosities, while the higher mass donors may become ultraluminous
for a few million years. For larger orbital periods, RLOF starts while
the donor is on the giant branch, which can lead to very high X-ray
luminosities for a short while. Intermediate mass black holes with a
lower mass donor may contribute to the emission of gravitational waves
in the LISA band and simultaneously be visible as bright X-ray
sources.

\section{The formation and evolution of an intermediate-mass black holes with 
         a stellar companion}\label{Sect:FandE}

\subsection{The formation}\label{Sect:formation}

Several theoretical models exist for producing black holes of
intermediate mass, (between $\sim10^2-10^4 M_\odot$; Portegies Zwart
\& McMillian 2002; Miller \& Hamilton 2002; Madau \& Rees
2001).\nocite{2002ApJ...576..899P} \nocite{2002MNRAS.330..232M}
\nocite{2001ApJ...551L..27M}

Young and dense star clusters may experience an evolutionary phase in
which the stellar density in the core becomes so high that physical
collisions between stars become frequent (Quinland \& Shapiro 1987;
Portegies Zwart et al 1999; Rasio, Freitag \& G\"urkan,
2003). \nocite{1987ApJ...321..199Q} \nocite{1999A&A...348..117P}
\nocite{2003astro.ph..8449G} Portegies Zwart \& McMillan
(2002)\nocite{2002ApJ...576..899P} derive a mass accumulation rate due
to collisions of $\Mdot \simeq 10^{-4} M_{\rm cl}/\trlx$, with $M_{\rm
cl}$ the mass of the star cluster and \trlx\, its half-mass relaxation
time. The requirement for this mass coagulation is that the initial
relaxation time of the cluster is less than $\sim 30$\,Myr, otherwise
the most massive stars which drive the collision runaway experience
supernovae before the star cluster experiences core collapse.  If the
cluster is born with sufficiently concentrated, a longer relaxation
time up to about 100\,Myr can still lead to a phase of collision
runaway (Portegies Zwart et al 2004).  The final mass of such a
episode of coagulation is about 0.1 percent of the initial cluster
mass Portegies Zwart \& McMillan (2002, see also G\"urkan et al 2004).

Following this scenario, intermediate mass black holes populate the
central regions of massive and dense star clusters. But the presence
of an intermediate mass black hole is no guarantee for X-ray
emission. The ambient gas density in such clusters is generally rather
low and is insufficient to power an ultraluminous X-ray source.

Another possibility would be the tidal disruption of a main sequence
star that smashes into the black hole or grazes it near the tidal
radius.  As a result, the star may be completely destroyed and
accreted by the black hole.  Tidal disruption with a supermassive
black hole leads to short flares (lasting about a year, Rees,
1988\nocite{1988Natur.333..523R}; Ulmer,
1999\nocite{1999ApJ...514..180U}; Ayal, Livio \& Piran,
2000\nocite{2000ApJ...545..772A}).  For lower mass black holes the
duration of accretion will longer inversely proportional with the mass
of the compact object. From this point of view, such a scenario is
consistent with the properties of ULX, but the event rate must be
impractically high to explain the number of ultraluminous X-ray
sources which have been observed, like M82-X1.

A more attractive model was recently proposed by Hopman, Portegies
Zwart \& Alexander (2004). They argue that in a young and dense
cluster a star may be captured by the intermediate mass black hole in
a tidal event. Further tidal interaction between the black hole and
the captured star then circularize the orbit.  They further assume
that the captured star is on the main-sequence, but the same
argumentation can be made for evolved stars.  As a orbit has been
fully circularized the captured star under-fills its Roche-lobe only
slightly (Hopman et al 2004). During the remainder of the
main-sequence lifetime of the captured star it grows in size by about
a factor of two and gravitational wave radiation reduces the orbital
separation. Ultimately the star fills its Roche lobe and mass transfer
to the intermediate mass black hole starts.

An alternative scenario for the formation of a binary is by the
dynamical capture of a main-sequence star by the intermediate mass
black hole. Such events will typically result in rather wide (100--
1000 day) binaries (Baumgardt et al.\, 2004). The companion to the
intermediate mass black hole is then be among the most massive stars
in the cluster. At young age ($\aplt 30$\,Myr) the companion and may
well be a massive ($\apgt 10$\,\msun) main-sequence star, but for
older cluster the companion is more likely to be another black hole.

The star clusters in which either tidal or dynamical capture can take
place are very young ($\aplt 10$\,Myr), extremely dense $\rho \apgt
10^5$\msun\,pc$^{-3}$ and rich $N \apgt 10^5$ stars (Portegies Zwart
et al.\, 2004). The relatively recently identified class of young
dense clusters (sometimes pronounced as YoDeC) tend to populate active
star forming regions in interacting galaxies, like M82 and the
Antennae (Zezas \& Fabbiano 2002). Hopman et at. (2004) argue that in
about half of the star clusters with are able to produce an
intermediate mass black hole via collision runaway, the back hole is
able to capture a companion star in a sufficiently short orbital
period binary that Roche-lobe overflow starts some time on the main
sequence.  Together with the possibility of dynamical capture, wide
binaries with an intermediate mass black hole may be comparatively
common as short period binaries.  We do not further consider the
details of dynamical or tidal capture, nor do we assess the
probability of forming the type of binary systems discussed in this
paper, as we mainly concentrate on the phase in the lifetime of the
intermediate mass black hole at which it is visible in X-rays or as a
gravitational wave source.

\subsection{The evolution of a binary with an intermediate mass 
            black hole}\label{Sect:binev}

Here we assume that an intermedate mass black hole was formed and has
acquired a stellar companion via one of the mechanisms discussed
above.  Mass transfer from a lower mass (secondary) star to a much
more massive (primary) star is driven by the expansion of the donor
and the loss of angular momentum from the binary system. In our
evolutionary model, we initialize the binary systems with a black hole
mass (typically 1000\,\msun) and a stellar mass secondary of
2\,\msun\, 5, 10 or 15\,\msun with some selected orbital period at
birth ($P_{\rm start}$). Apart from the previous model description in
\S\,\ref{Sect:formation} we do not further discuss how such parameters
are obtained.

The evolution of each selected secondary mass is calculated with an
initial orbital period such that Roche-lobe overflow starts at
zero-age, which could be representative if the system was formed via
tidal caputre.  This choice seems somewhat arbitrary, as it leaves no
time for the IMBH to form and to acquire Roche-lobe contact; to first
order however, this is a good assumption as it takes less than about
three million years to form the IMBH (Portegies Zwart et al 2004), and
the capture of a companion star followed by tidal secularization takes
only a few million years (Hopman et al,
2004).\nocite{2004ApJ...604L.101H} For our 2\,\msun\, donors this time
lag is negligible, while for the 15\,\msun\, donor it is roughly half
the main-sequence lifetime.  We still argue, however, that the exact
moment of RLOF has little effect for the further evolution of the
binary system, as long as it starts somewhere on the
main-sequence. Some binaries start Roche-lobe overflow at a later
evolutionary stage, on the giant or super giant branches.  Such larger
initial orbital period could be representative in the case that the
binary was formed by a dynamical capture. An overview of the selected
initial conditions is presented in Tab.\,\ref{Tab:models}.

Once initialized, we evolve the binary through various stages using
the binary evolution code of Eggleton (see Pols et
al. 1995\nocite{1995MNRAS.274..964P} and references therein for a
description), assuming a population I chemical composition (Y = 0.98,
Z = 0.02) and mixing-length parameter of $\alpha = 2.0$. Convective
overshooting is taken into account using an overshooting constant
$\delta_{\rm ov} = 0.12$ (Pols et
al. 1998)\nocite{1998MNRAS.298..525P}.

The orbital evolution of the system is affected by the emission of
gravitational waves (Landau \& Lifshitz 1958),\nocite{1958.book.....L}
Roche-lobe overflow by non-conservative mass transfer or mass loss via
a wind (Soberman et al. 1997).\nocite{1997A&A...327..620S} The
evolution of the orbital separation, $a$, can then be computed by the
sum of three terms
        \begin{eqnarray}
          \frac{\dot{a}}{a} & = & - \frac{64 \, G^{3}}{5 \, c^{5}} 
          \frac{m_{{\rm d}} \, \Mbh \, m_{{\rm T}}}{a^{4}} + 
          \frac{2 \, [\beta \, q^{2} - q + (\gamma - 1)]}{1 + q} 
          \frac{\dot{m}_{{\rm d}}}{m_{{\rm d}}} \nonumber \\ 
                            &   & - 2 \frac{\dot{m}_{{\rm BH}}}{m_{{\rm BH}}} + 
          \frac{\dot{m}_{{\rm T}}}{m_{{\rm T}}}.
          \label{doublens:eq:orbit}
        \end{eqnarray}
Here $G$ is the constant of gravity, $c$ is the speed of light in
vacuum, $m_{{\rm d}}$ and $m_{{\rm BH}}$ are the masses of the donor
star and the black hole, respectively, $m_{{\rm T}} = m_{{\rm d}} +
m_{{\rm BH}}$ and $q = m_{{\rm d}} / m_{{\rm BH}}$.  The parameter
$\gamma$ gives the fraction of mass lost from the donor star in the
form of fast isotropic wind (de Jager et
al. 1988)\nocite{1988A&AS...72..259D} and $\beta$ is the fraction of
the wind mass accreted by the black hole. During the detached phase,
we assume that $\gamma = 1$ and that the black hole does not accrete
mass from the stellar wind ($\beta = 0$). Matter leaves the system
carrying the specific angular momentum of the donor. During
mass-transfer we assume that the black hole accretes matter up to its
Eddington limit (see e.g.\, King 2000)\nocite{2000MNRAS.312L..39K} as
        \begin{eqnarray}
          \dot{M}_{{\rm Edd}} = 10^{-8} m_{{\rm BH}} {\rm M_\odot / yr}
        \end{eqnarray}
The remaining mass is lost from the system with the specific angular
momentum of the black hole. A more sophisticated Eddington mass
accretion is given by Podsiadlowski et al. (2003, see their eq.\,9),
which depends on the hydrogen abundance of the donor and the initial
and present mass of the black hole.  Because in our calculations
$M_{{\rm d}} \ll M_{{\rm BH}}$, throughout the evolution the present
mass of the black hole does not differ much from its initial mass, and
therefore our adopted Eddington limit is about a factor $\sim$ 4.3
lower than that of Podsiadlowski et al. (2003). As the consequence,
the black holes in our case accrete less mass and the orbital period
expands more strongly.

Table\,\ref{Tab:models} gives an overview of the initial conditions
and results of our calculations.  In Fig. \ref{fig:tMdotEdd_ZAMS} we
present the evolution of mass-transfer rate for our calculations where
mass transfer started at the zero-age main sequence.

\subsubsection{Case A mass transfer}

Binaries which start Roche-lobe contact at the ZAMS undergo case A
mass transfer. This prolongs their main-sequence lifetime
considerably.  The most peculiar evolution in our sample is arguably
that of the 2\,\msun\, donor.  After transferring about 1~\msun\, the
star starts to develop a convective envelope which initiates chemical
mixing, which again leads to magnetic activity. As a result orbital
evolution will become dominated by magnetic braking. This mechanism
ceases when the whole star becomes convective with a homogeneous
composition of $X = 0.66$.  Some time later, the central temperature
drops below the hydrogen burning limit and thermonuclear fusion stops:
this happens at about $t=630$\,Myr. At that point the system remains
detached for about 8\,Myr before it undergoes another phase of mass
transfer.  At $t=1.2$\,Gyr the code breaks while the donor still fills
its Roche lobe; at this point $T_{\rm eff} =$~840~K and $M =$ 0.011
\msun\, (about 10 $m_{\rm Jup}$), which can be considered a brown
dwarf. Under the influence of gravitational-wave radiation alone, the
donor spirals in toward the black hole until they merge after about 74
Myr (Peters 1964).\nocite{peters64}

\begin{figure}
\psfig{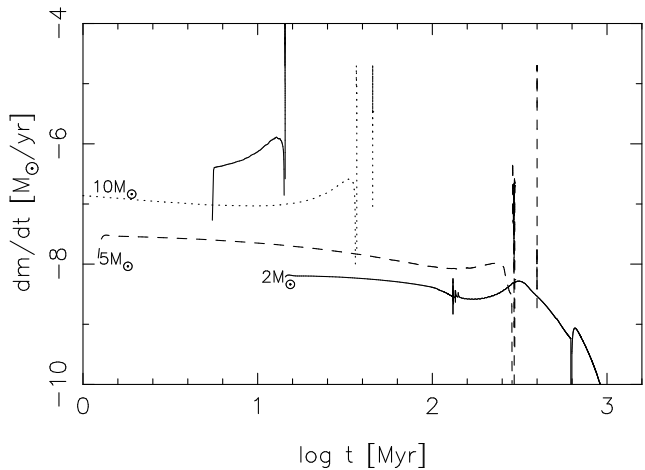}
\caption{Evolution of mass-transfer rate onto a 1000\,\msun\, black
hole from a donor with an initial mass of 2\,\msun\, (lower solid), 5
(dashes), 10 (dots) and 15\,\msun\, (upper solid), and initial orbital
period of 0.5\,days, 0.6, 0.8 and 1.0 day, respectively.
\label{fig:tMdotEdd_ZAMS}
}
\end{figure}

For the more massive stars, case A mass transfer is followed by mass
transfer when the star ascends the giant branch during hydrogen shell
burning (case AB), lasting for 8.3 Myr (for $\mdon = 5$ \msun),
0.26\,Myr (for 10\,\msun) and 0.05\,Myr (for $\mdon = 15$ \msun). The
mass-transfer rate increases dramatically in this phase to 5.2 $\times
10^{-7}$, $3.7 \times 10^{-5}$ and 2.3 $\times 10^{-4}$ \msun/yr for
the 5\,\msun, 10\,\msun\, and 15\,\msun\, donor, respectively; as
shown by the first spike in the evolution in
Fig.\,\ref{fig:tMdotEdd_ZAMS}. These mass transfer rates result in
X-ray luminosities in excess of $10^{41}$ erg/s.

The 10 \msun\ donor experiencing case A mass transfer ends as a helium
star. This star experiences helium shell burning, lasting for
about 40\,000 years, during which mass transfer starts again at a rate
of $3 \times 10^{-6}$ \msun/yr, regardless of the black-hole
mass. These phases of mass transfer are shown by the second spikes in
Figs.\,\ref{fig:tMdotEdd_ZAMS} for 5\,\msun\, and 10\,\msun\,
donors. Though seemingly unresolved, some of these spikes last for
several million years and are well resolved in our simulations (see
also Tab.\,\ref{Tab:models}).

\subsubsection{Case B and C mass transfer}

Binaries initialized with a larger orbital period experience mass
transfer when the donor is evolved to a giant (case
B)\footnote{Although they are both initiated on the giant branch, we
distinguish case B mass transfer from case AB, as case B which is
preceded by case A mass transfer.  Consequently, we use case BC as
case C which is preceded by case B mass transfer.} or supergiant
(case C). The case B systems remain detached for only 0.29\,Myr (5
\msun) -- 0.05\,Myr (10 \msun) at mass-transfer rate of 2.4 $\times
10^{-5}$ (5 \msun) -- 7 $\times 10^{-4}$ \msun/yr (10 \msun).

The 10 \msun\ donors which experience case B mass transfer ascends the
super giant branch at a later instant, leading to case BC mass
transfer. The mass-transfer rate during this stage is of the same
order as in the preceding phase.  The evolution of the 10 \msun\ donor
in a $\sim 800$ day orbit around a 1000 \msun\ black hole is quite
peculiar; mass transfer, in this system, ensues in a relatively
advanced stage (late case B) such that only about 0.5 \msun\ is
transferred before the hydrogen shell burning is terminated, causing
the star to shrinks inside its Roche lobe. Another 2 \msun\ is
transferred during case BC mass transfer before the donor turns into a
helium star, which experiences helium shell burning, during which it
expands quickly to overfill its Roche lobe, initiating a short phases
of high ($10^{-5}$ \msun/yr) mass transfer. Also the 5 \msun\, donor
ultimately turns in a degenerate CO core and experiences several
carbon flashes during its remaining lifetime during which the mass
transfer rate varies between $\sim 10^{-7}$ \msun/yr and $10^{-5}$
\msun/yr.

Only one of our simulations experienced case C mass transfer, which
occurred for the 5 \msun\ donor with a 600 day orbital period.  This
resulted in a short phase of mass-transfer at a rates of 1.4 $\times
10^{-5}$ \msun/yr.

Only the binaries with a 10 \msun\ donor for case B, and 15 \msun\
for case A, leave neutron stars at the end of their evolution; all
other evolutions leave carbon-oxygen white dwarfs.

\begin{table*}
\caption[]{Initial conditions and results of our detailed binary
evolution calculations. The first three columns give the initial
conditions: black-hole mass, companion mass (both in \msun) and
orbital period (days). The subsequent three columns give the time at
which the calculation is stopped (in Myr), the mass of the donor at
this moment (\msun) and the orbital period (days). The following 4
columns give the duration of RLOF, and the duration for which 
mass-transfer exceeds the Eddington limit for a 1\,\msun\,
accreting compact object, 10 times and 100 times the Eddington
rate. The last column gives an estimate for the duration (in Myr) over
which the binary is visible as a gravitational wave source in the LISA
band assuming a distance of about 10\.kpc. Time resolution is about
0.05\,Myr.  }
\begin{flushleft}
\begin{tabular}{lll|lll|rrrrrcc} \hline
$m_{\rm BH}$    & $m_{\rm d}$ & $P_{\rm start}$  
             & $t_{\rm end}$ & $m_{\rm end}$ & $P_{\rm end}$ 
             & $\Delta t_{\rm \dot{m}>0}$ 
             & $\Delta t_{\rm \dot{m}>\dot{m}_{\rm Edd}}$ 
             & $\Delta t_{\rm \dot{m}>10\dot{m}_{\rm Edd}}$ 
             & $\Delta t_{\rm \dot{m}>100\dot{m}_{\rm Edd}}$ 
             & $\Delta t_{\rm GWR}$ 
             \\
\multicolumn{2}{c}{--- [\msun] ---} & [days] 
       & [Myr] & [\msun] & [days]
       & \multicolumn{5}{c}{--- [Myr] ---}  
       \\
\hline
1000  &  2&   0.5& 1200& 0.01& 0.09& 1189 &  0.0  &  0     & 0   & 1200 \\
1000  &  2&  50  & 1248& 1.71& 87.4&  2.0 &  2.0  &  0     & 0   & 0    \\
1000  &  5&   0.6&  398& 0.61&  161&  294 & 42.7  &  4.40  & 0.55& 130  \\
1000  &  5&   6.0&  123& 0.96&  846&  0.8 &  0.80 &  0.75  & 0.55& 0    \\
1000  &  5&  60  &  123& 1.18& 4510&  0.70&  0.65 &  0.65  & 0.60& 0    \\
1000  &  5& 600  &  123& 2.31& 6430&  0.85&  0.85 &  0.85  & 0.20& 0    \\
100   & 10&   0.8& 46.3& 1.04&  976&  36.9& 36.9  & 13.4   & 0.80& 32   \\
500   & 10&   0.8& 45.0& 1.05&  612&  36.1& 36.1  & 12.7   & 0.80& 23   \\
1000  & 10&   0.8& 45.6& 1.05&  606&  36.6& 36.5  & 12.3   & 0.80& 15   \\
2000  & 10&   0.8& 47.1& 1.03&  589&  37.6& 37.6  & 12.5   & 0.80& 3    \\
1000  & 10&   8.0& 27.2& 2.57&  477&  0.05&  0.05 &  0.05  & 0.05& 0    \\
1000  & 10&  80  & 26.9& 3.57& 1795&  0.05&  0.05 &  0.05  & 0.05& 0    \\
1000  & 10& 800  & 26.9& 7.22& 2166&  0.05&  0.05 &  0.05  & 0.05& 0    \\
1000  & 15&   1.0& 16.2& 3.56&  100&   8.9&  8.9  &  8.7   & 0.05& 12   \\
\hline
\end{tabular} 
\end{flushleft}
\label{Tab:models} 
\end{table*}

The durations of mass transfer are presented in columns 7--10 of
Tab.\,\ref{Tab:models}. Only if Roche-lobe overflow occurs at an early
stage, near the ZAMS, can mass transfer be sustained at high rates for
a long time. High accretion rates, however are only achieved on
the giant (case B) and supergiant (case C) branches. We also find that
mass transfer caused by helium shell burning can result in mass
transfer rates as high as $10^{-5}$ \msun/yr, which is sufficient to
power an $10^{40}$\, erg/s X-ray source, but only for a short while.

\section{Observational characteristics}\label{Sect:Observations}

In this section we apply and discuss an empirical observational model
based on our understanding of low-mass X-ray binaries in the Galaxy,
but then applied to a system where the donor is a 2--15\,\msun\, star
and the accretor is an intermediate mass black hole. We then use the
results of our binary evolution calculations as input for the
empirical model. In Appendix\,\ref{Sect:Appendix} we present the
empirical evidence on which this model is based.

The binary evolution simulations provide us with the donor mass,
accretor mass, orbital period and a secular mass accretion rate, all
as a function of time. We use these information to calculate the X-ray
luminosity of the simulated binaries (below); the gravitational wave
signatures are discussed in \S\,\ref{Sect:GWR}.

Our binary evolution calculations provide secular accretion rates,
rather than instantaneously observable ones.  This difference can be
quite important, considering that a large fraction of the stallar mass
accreting black holes are transients.  Since the theoretical
understanding of disk instabilities is not presently well enough
established to make quantitative predictions, we develop an {\it ad
hoc} model based on phenomenology (discussed in
Appendix\,\ref{Sect:Appendix}). The parameters we require in order to
compare our simulations with observations are the recurrence time
($t_{\rm r}$) and the decay time ($t_{\rm d}$), via which we calculate
the time $t_{\rm br}$ that the source is brighter than a certain
limiting X-ray luminosity $L_{\rm br}$.  We then compute the duty
cycle ($\tau_{\rm duty} \equiv t_{\rm br}/t_{\rm r}$).

After applying the empirical model for X-ray emission, we study the
observational characteristics of these sources.  We require that to
recognize the source as a ULX it has to be brighter than $L_{\rm br}$,
for which we adopt $L_{\rm br} = 10^{40}$ erg/s.  This is done in the
following way: First we calculate the total energy from a fit to the
empirical data in Fig.\,\ref{fig:Porb_E} and the decay time from
Fig.\,\ref{fig:Ltr}.  The recurrence time is subsequently a function
of the total energy $E$ and the rate of mass transfer via, $t_{\rm r}
= E/(0.1\dot{M}c^2)$.  The bright time is subsequently calculated by
integrating the luminosity curve, for which we assume an exponential
decaying function (see Appendix\,\ref{Sect:Appendix}) from the start
to the moment it drops below $L_{\rm br}$. This results in $t_{\rm br}
= t_{\rm d} - \log( L_{\rm br}/2L_{\rm Edd})$, where we adopt the
factor two in the denominator as seems reasonable from
Fig.\,\ref{fig:Ltr}. The duty cycle is subsequently determined as
$\tau_{\rm duty} = t_{\rm br}/t_{\rm r}$.

\begin{figure}
\psfig{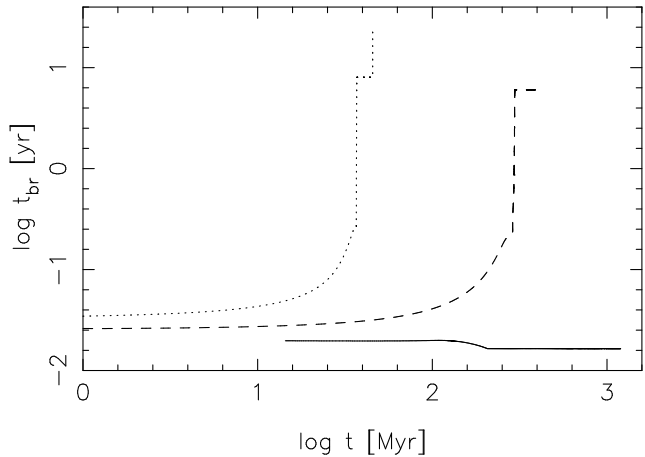} 
\caption{Time spend (on average) per duty cycle as a
$L_x>10^{40}$\,erg/s x-ray source for a binary with a 1000\,\msun\,
black hole and a 10\,\msun\, 5\,\msun\, and 2\,\msun\, donor which
starts Roche-lobe overflow at birth.
\label{fig:bright_time}
}
\end{figure}

\begin{figure}
\psfig{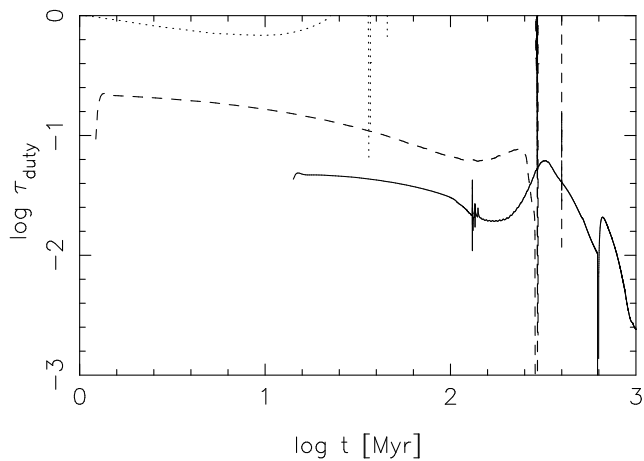} 
\caption{Duty cycle (for $L_X > 10^{40}$\,erg/s) for the three
binaries with starts RLOF at birth with a 2\,\msun\, donor (solid) a
5\,\msun\, (dahses) and a 10\,\msun\, (dotted line) donor.  A duty
cycle $\tau_{\rm duty} > 1$ indicates that the source is persistent.
\label{fig:duty_cycle}
}
\end{figure}

In figures\,\ref{fig:bright_time} and\,\ref{fig:duty_cycle} we present
the resulting bright time ($t_{\rm br}$) and duty cycle ($\tau_{\rm
duty}$) for a selection of binaries with 2\,\msun, 5\,\msun and
10\,\msun\, donors which start RLOF at the ZAMS.  The binaries which
start at wider orbits and the simulations for a 15\,\msun\, donor
which starts RLOF at the ZAMS are all persistent X-ray source with
luminosities well above $L_{\rm br}$.  The binaries with a relatively
wide initial orbit, however, remain bright for only a short while.

Our empirical prescription is supported by the qualitative
similarities with the proposed limit for transient and persistent
sources of Dubus et al. (1999), which yields that binaries with
accretion rates exceeding
\begin{equation}
	\mdot_{\rm crit} \simeq 8.5 \times 10^{-11} \msun {\rm yr}^{-1}
	\left( \Mbh \over 10^3\msun \right)^{0.5}
	\left( \mdon \over \msun \right)^{-0.2} 
	\left( P \over {\rm days} \right)^{1.4}
\label{Eq:dubus1999}\end{equation}
are persistent X-ray emitters.  With this prescription, lower mass
donors ($\mdon \aplt 10$\,\msun) show transient behavior.

\section{Discussion}\label{Sect:discussion}

We have analyzed the results of evolutionary calculations in which a
donor star of 2\,\msun\, -- 15\,\msun\, transfers mass to a 100 ---
2000\,\msun\, black hole.  In this section we discuss our simulation
results with respect to the observed bright X-ray source M82-X1 in the
starburst galaxy M82.  This source is one of the best candidate
intermediate mass black holes, and we argue that such a black hole may
produce X-rays as it receives mass from a companions star.

Binaries which start RLOF at the ZAMS remain bright only for a few
weeks continuously. Near the end of the main sequence the bright time
increases to several years only for the 5\,\msun\, and 10\,\msun\,
donors. Our calculation for a 2\,\msun\, donor only experiences a
short (few days) bright ($L_x>L_{\rm br}$) flash every few
months. Since the ultraluminous X-ray source M82-X1 was observed for
an extended period (over 20 years) we conclude that it is unlikely to
host a low mass ($\aplt 2$\,\msun) donor. Of course, it could be that
the X-ray source was caught near peak luminosity each time it was
observed, in which case a lower mass donor would prove satisfactory,
but this is unlikely. Our simulations for the 5\,\msun\, and
10\,\msun\, donor on the main-sequence exhibit similar episodes of
bright time but with shorter duty cycle, making these sources still
rather unlikely candidates for the ULX in M82.  However, near the end
of the main sequence, $t_{\rm br}$ increases considerable to several
years. The binaries with a 10\,\msun\, donor then become steady
sources, making these suitable candidates for the ULX in M82. The
15\,\msun\, donor which starts RLOF at the ZAMS is also persistently
bright.

Strictly speaking, the systems with an evolved donor are transients,
but for as long as RLOF is sustained the bright time and recurrence
time become so long that these sources could rather easily be
identified as persistent ultraluminous X-ray emitters.  And though
these sources are rather short lived they should not be excluded as
suitable candidates (see also Kalogera et al 2004).


Based on the above analysis and the figures\,\ref{fig:bright_time} and
\,\ref{fig:duty_cycle} we conclude that within the limitations of our
model the most likely configuration for the bright X-ray source M82-X1
is a 1000\,\msun\, with a 10-15\,\msun\, donor, probably near the end
of its main-sequence lifetime. Note that for the star cluster MGG-11
this would also be consistent with the observed age of the cluster,
being 7-12\,Myr. The turn-off age of a 15\,\msun\, star
is about 11.5\,Myr.  In a phase of mass transfer the stellar
lifetime can be extended considerably possibly allowing even more
massive donors (see Section\,\ref{Sect:binev}).

\subsection{Emission of gravitational waves}\label{Sect:GWR}

\begin{figure}
\psfig{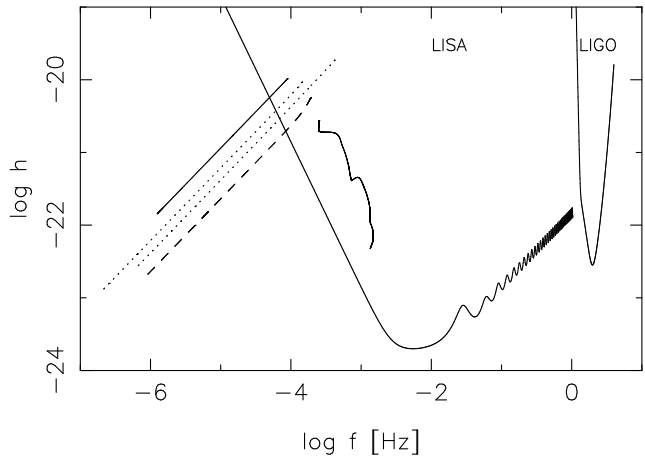} 
\caption{A selected sample of binaries in the gravitational wave
strain and frequency domain with the LISA and LIGO noise curves
over-plotted.  We assumed a distance to the source of 10kpc.  The
lower solid curve to the right is for a 2\,\msun\, donor, the dashes
for a 5\,\msun\, donor, the dotted curves are for a 10\,\msun\, donor
with a 100\,\msun\, black hole (lower dotted curve) and 1000\,\msun\,
black hole (upper dotted curve). The upper solid curve is for a
15\,\msun\, donor. Except for the lower dotted curve are all
calculations for a 1000\,\msun\, donor. Notice the enormous difference
for the 2\,\msun\, donor, which is caused by the perturbed evolution
of the donor (see Sect.\,\ref{Sect:binev}).
\label{fig:tMdotEdd_GWR}
}
\end{figure}

A binary emits gravitational waves, and as long as the stars are far
apart we can assume that they behave as point masses. In the case we
consider here, the donor star fills its Roche-lobe and can hardly be
described as a point mass.  The amplitude of the gravitational wave
signal is given by (Evens et al.\, 1987)
\begin{equation}\label{eq:h}
  h \simeq 6 \times 10^{-23} \left( \frac{\mathcal{M}}{\msun} \right)^{5/3}
\! \! \! \left( \frac{P_{\rm orb}}{\rm 1\, {\rm days}} \right)^{-2/3} \!
\! \!\left( \frac{d}{\rm 1 {\rm kpc}} \right)^{-1}.
\end{equation}
Here ${\cal M} = (\Mbh\mdon)^{3/5}/(\Mbh+\mdon)$ is the chirp mass.
The gravitational wave frequency is twice the orbital frequency.  The
resulting gravitational wave signal is presented in
Fig.\,\ref{fig:tMdotEdd_GWR}, for binaries which started Roche-lobe
overflow on the zero-age main-sequence. Binaries which do not start
RLOF during the main-sequence phase of the donor will not become
detectable gravitational wave sources.

Figure\,\ref{fig:tMdotEdd_GWR} presents the evolution of the
gravitational wave signal for several systems which start RLOF at the
ZAMS.  The last column of Tab.\,\ref{Tab:models} presents the time
that the binary is visible above the noise curve of the LISA
gravitational wave detector.

Assuming a standard distance of 10\,kpc, the binaries which start mass
transfer at birth emit gravitational waves at a strain of about $\log
h = -20.2$ (almost independent of the donor mass). The frequency of
the gravitational waves is barely detectable for the {\em LISA} space
based antennae (see Figure\,\ref{fig:tMdotEdd_GWR}). During mass
transfer the binary moves out of the detectable frequency band, as the
orbital period increases.  Once the donors has turned into a white
dwarf, as happens for the 5\,\msun\, and 10\,\msun\, donors, the
emission of gravitational waves brings the two stars black into the
relatively high frequency regime and it becomes detectable again for
the {\em LISA} antennae. This process, however takes far longer than a
Hubble time, unless the orbit has an eccentricity $e > 0.97$. Such
high eccentricities, after the phase of mass transfer, can only be
achieved in the cases where the initial donor collapses to a compact
object in a supernova explosion, or if the binary is perturbed by
external influences. We conclude that the here discussed class of
binaries with relatively massive donors $\apgt 5$\,\msun\, are
probably not important sources of gravitational waves.

The binary in which a 2\,\msun\, main-sequence star starts to transfer
to a 1000\,\msun\, black hole, however, remains visible as a bright
source of gravitational waves for its entire lifetime (see
Figure\,\ref{fig:tMdotEdd_GWR}). The reason for this striking result
is the curious evolution of the donor (see sect.\,\ref{Sect:binev}).
Mass transfer in this evolutionary stage is rather slow causing the
X-ray source to be transient. We argue that such a transient could
result in an interesting synchronous detection of X-rays and
gravitational waves. The known ULXs are at Mpc distance and unlikely
to be seen with LISA.

\section{Summary}

We performed detailed simulations of the evolution of close binaries
with an intermediate mass (100--1000\,\msun) black hole and a stellar
companion.  If the initial orbital period is of the order of a day,
the main-sequence star starts filling its Roche lobe before the
terminal age. In this case the binary remains in contact for it's
remaining main-sequence lifetime, which is considerably extended
compared to the main-sequence lifetime of an isolated star of the same
mass.

Relatively low mass ($\aplt 5$\,\msun) donors which start mass transfer
at zero age are likely to be transient, with high
($>10^{40}$\,erg/s) peak luminosity. The recurrence time for these
sources, however, is generally rather long (several weeks) and the
duration of the bright phase is short (few days).  More massive donors
are bright for a longer time and have shorter recurrence time.  The
sources become persistent for donor stars of about 10\,\msun.  Our
simulation with a 15\,\msun\, ZAMS donor resulted in a persistent
bright X-ray source for its entire main-sequence lifetime.

The duty cycle also increases near the end of the main-sequence. Stars
which start Roche-lobe overflow after the main-sequence phase remain
persistent for their mass transferring episode, which, for a
5\,\msun\, donor lasts for $\sim 9$ million years.

Based on these findings, and within the limitations of our analysis
and assumptions, we argue that low mass ($\aplt 5$\,\msun) donors and
donors near the zero-age main-sequence are unlikely candidates for the
ultraluminous X-ray source M82-X1. A 10--15\,\msun\, donor near the
terminal-age main sequence or on the early giant branch produces
recurrence times and luminosities consistent with the observed ultra
luminous x-ray source in M82.

An interesting possibility is provided by a $\sim 2$\,\msun\, donor
which starts to transfer mass to an intermediate mass black hole at
birth. Such a binary is likely to be visible as a bright transient
X-ray source and simultaneously as gravitational wave source in the
{\em LISA} band to a distance of several Mpc.

\section*{Acknowledgments}
We are grateful to Chris Belczynski, Clovis Hopman, Peter Jonker,
Vicky Kalogera, Emmi Meyer-Hofmeister, Onno Pols, Ed van den Heuvel,
Lev Yungelson, Andrzej Zdziarski, Steve Zepf and Andreas Zezas for
discussions and for providing useful data. This work was supported by
the Royal Netherlands Academy of Sciences (KNAW), Netherlands Research
School for Astronomy (NOVA) and the Dutch Organization for Scientific
Research (NWO) for the Spinoza Grant 08-0 to E.~P.~J. van den Heuvel
and a Talent Fellowship.  We are grateful to Northwestern University
for hospitality.

\appendix  
\section[]{Appendix: Empiric observational characteristics}\label{Sect:Appendix}
In this appendix we discuss the characteristics of low-mass black hole
X-ray binaries in the Galaxy, in order to compare them with the X-ray
binaries we have been simulating in the main body of the paper. The
main reasons to use the Galactic population of soft X-ray transients
is because of the similarities in the geometry and composition of
these systems with accreting intermediate mass black holes.

\subsection{Some characteristics of the Galaxtic X-ray binaries}

Most accreting black holes in the Milky Way are transient systems.  In
fact, only Cygnus X-1 is a persistent emitter among the dynamically
confirmed black hole candidates (see McClintock \& Remillard 2003 for
a review), and the only other persistent emitters considered to be
plausible black hole candidates are 4U~1957+115, SS 433 and Cygnus X-3, along
with the Large Magellanic Cloud sources LMC X-1 and LMC X-3.  Of
these, only 4U~1957+115 has a low mass (i.e. $<10M_\odot$) donor star,
and the case for its being a black hole is by far the weakest (see
Nowak \& Wilms 1999).  On the other hand, there are many persistent
neutron star accretors.  Since standard accretion disk theory suggests
that disk instabilities occur due to changes in the disk's hydrogen
ionization state, we expect stable disks only when the disk is hot
enough for its outer edge to be above the ionization temperature of
hydrogen.  Given that more massive accretors have lower inner disk
temperatures (Shakura \& Sunyaev 1973) and larger orbital separations
(and hence bigger disks) for the same orbital period (Kepler 1609), it
is then not too surprising that neutron stars can more easily be
persistent accretors than black holes.  It thus becomes rather likely
that intermediate mass black holes will be transient accretors (see
the discussion in Dubus et al. 1999).

The most comprehensive study to date of the outburst properties of
soft X-ray transients is that of Chen, Shrader \& Livio (1997,
hereafter CSL97).  This work compiled the outburst durations, total
fluencies, and peak luminosities for all the observed X-ray transient
outbursts in the literature at the time of its writing, considering
both optical and X-ray properties of the outbursts.  We will borrow a
few key results from that paper, while also making a few
simplifications, to allow for a prescription for converting secular
accretion rate into X-ray luminosities that is suitable for porting to
accreting intermediate mass black holes.  We will supplement these
data with some more recent work, including that by Garcia et
al. (2003) who tabulated the peak outbursts luminosities of several
X-ray transients, and other data sets which will be specified in more
detail.

The frequencies of outbursts are poorly constrained both by
observations and by theory.  A reasonable approximation, then, is to
assume that the outbursts we see are ``typical'', and hence that
roughly periodic, identical outbursts will be seen for sources, with
the periodicity and the properties of the outburst varying from source
to source.  To estimate the recurrence timescale ($t_{\rm r}$), we
compare the secular mass accretion rate from our binary evolution
simulations (\S\,\ref{Sect:binev}) with an estimate of the total
amount of mass accreted during an outburst based on the energy emitted
during the outburst, $E$, plus an estimate of the radiative
efficiency. We then apply the observational characteristics to our
binary evolution calculations, which is done in
\S\,\ref{Sect:Observations}.

The first assumption we make is that the outburst follow a fast-rise,
exponential decay (FRED) luminosity law as a function of time.  This
approximation is supported by the shape of the light curves for about
half the outbursts studied in Chen, Shrader \& Livio (1997).  The
other half, and in particular GRS~1915+105 the Galactic system most
analogous to a ULX, show rather ``flat-topped'' outburst light curves
(Chen, Shrader \& Livio 1997).

The next assumption is that the orbital period of a system is the key
parameter in determining the properties of its outbursts.  The radial
size of the accretion disk is likely to be the actual driver of the
outburst properties, but it is hard to measure.  Given simple scaling
laws, though, and the assumption that the compact accretor dominates
the mass of the binary system, which is here the case, it can be shown
that the mass content of an accretion disk depends primarily on the
orbital period, with a linear dependence on mass of the accretor.  The
outer radius of the disk, $r_{\rm disk}$ then scales with the orbital
separation, $a$, since $a\propto$$m_{BH}^{1/3}$; and the mass of the
disc is proportional to $r_{\rm disk}^3$; then the mass of the
accretion disk is proportional to the mass of the accretor.  If we
then scale all observational quantities in Eddington units, the mass
of the accretor has relatively weak effects on the properties of the
outbursts.

\begin{figure}
\psfig{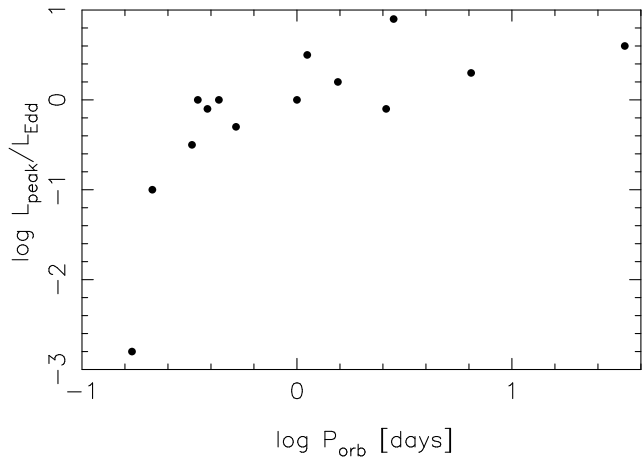}
\caption{The peak luminosity as fraction of the Eddington luminosity
for well studies X-ray binaries with a black hole (data from Garcia et
al., 2003).
\label{fig:Ltr}
}
\end{figure}

Fig.\,\ref{fig:Ltr} shows the peak luminosity at outbursts as a
function of the orbital period for several black hole X-ray binaries
(data from Garcia et al. 2003). There are relatively few systems, but
the systems in short orbital periods (i.e. less than about 10 hours)
fit approximately to the relation $L_{\rm peak}=2 L_{\rm Edd}\times
(P_{\rm orb}/10 {\rm hours})$, with the sources above 10 hours fitting
approximately to $L_{\rm peak} = 2 L_{\rm Edd}$, which is also
supported by theoretical studies, which suggests that the amount of
mass accumulated in the disc in an outburst cycle is a function of the
orbital period with a cutoff around 10 hours (see Meyer-Hofmeister \&
Meyer, 2000; Meyer-Hofmeister, 2004).  We note that the systems used
to make the correlations between outburst properties and orbital
periods appear to be prediminantly at most only mildly evolved, so
that the applications of these results to highly evolved systems
(i.e. giant donors) could be dangerous.

The two outliers are V4641 Sgr, at 67.6 hours orbital period, and
$L\sim8L_{\rm Edd}$ and XTE~J~1118+480, with $P_{\rm orb} \simeq
4.1$\, hours and $L_{\rm peak} \sim 0.001 L_{\rm Edd}$.  Since V4641
Sgr shows a radio jet with an apparent proper motion of more than 10
times the speed of light (Hjellming et al. 1999; Orosz et al. 2001)
its jet must be pointed very close to the line of sight, and if its
jet has a substantial X-ray spectral component, then this beamed
component may be responsible for the strong X-ray emission, in
contrast to what is expected for the other sources.  The other
outlier, XTE~J~1118+480, is one of the X-ray transients that never
reached a high luminosity state (see e.g. Brocksopp, Bandyopadhyay \&
Fender, 2004, for a discussion of such sources).  Since these low
luminosity systems are thought to be radiatively inefficient, it may
be that the instantaneous mass transfer rate did reach levels
comparable to the level of the correlation (indicating that the disk
outburst physics was essentially the same for this source as the
others).  Alternatively, since the outbursts of individual recurrent
X-ray transients are not all identical, it may simply be that we saw a
fainter than typical outburst of XTE~J~1118+480.  Given that the
primary goal of this paper is to develop a model for {\it bright}
X-ray binaries, this outlier is not particularly troublesome.

\begin{figure}
\psfig{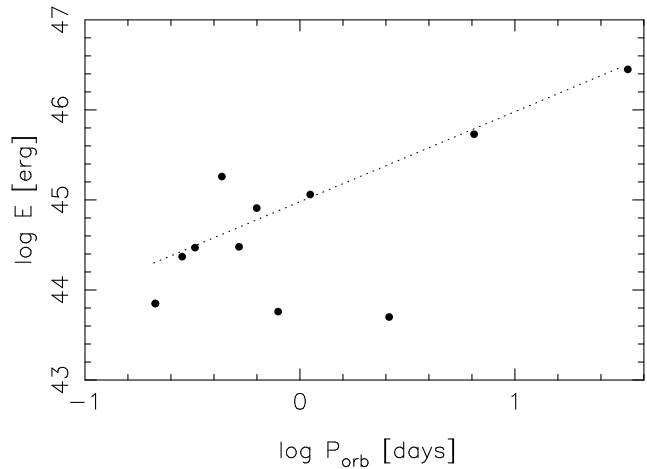}
\caption{The outburst energies for X-ray transients with well-measured
peak luminosities and orbital periods.  We can fit the relation with
$\log (E/{\rm erg}) = 45 + \log(P_{\rm orb}/{\rm days})$ (data from Garcia et
al., 2003).
\label{fig:Porb_E}
}
\end{figure}

Next, we consider the outburst durations.  Taking the data from CSL97
for black hole binaries' outburst energies, we find a similar relation
with orbital period, that $\log E= 45 \log(P_{\rm orb}/10 {\rm
hours})$, as shown in Fig.\,\ref{fig:Porb_E}.  We note that the strong
outlier is GRO~J~1655-40, which has shown multiple outbursts with a
short recurrence time, highlighting the limitations of the ``periodic
outburst'' hypothesis, which we use here as a simplification.  We also
note that the highest energy, longest period system, GRS~1915+105 has
been in outburst continuously since about 1992, but the data from Chen
et al. (1997) obviously covers only a fraction of that timespan.
Using the value from Chen et al. (1997), it falls a factor of about 3
below the relation we propose between orbital period and outburst
energy.  Combining the two empirical relations from the data, with the
added assumption of a ``FRED'' lightcurve shape, we then find that the
exponential decay timescale $t_{\rm d}$ for an outburst should be
approximately
\begin{equation}
  t_{\rm d} = 6 {\rm days} \times 
  \min \left(1, {P_{\rm orb} \over 10 {\rm hours}}\right).
\end{equation}
We emphasize that the prescription presented here is a crude
approximation to the poorly understood phenomenology of transient
disks; however, we feel it is a better approximation that the standard
ones made in converting secular accretion rates into instantaneous
mass accretion rates in the X-ray producing regions of the accretion
flow. The theoretical model recently proposed by Meyer-Hofmeister
(2004) for outbursts also predict a correlation between orbital period
and peak output luminosity.  We note especially that the FRED shape
may be a poor approximation for the ULXs, considering that
GRS~1915+105, the Galactic system most analogous to the ULXs, has
shown a rather ``flat-topped'' outburst light curve.

Based on this phenomenological analysis we conclude that the peak
luminosity for a $\sim 1000$\,\msun black hole binaries easily exceed
$10^{41}$ erg/s, but these high luminosities are generally sustained
for a short while only (see also \S\,\ref{Sect:Observations}).

\subsection{Spectral State Phenomenology: Going from Narrow Bandpass Luminosity to Mass Accretion Rate}

Next, we provide a brief review of spectral states of X-ray binaries,
which will be used to justify bolometric corrections to the data.
Between three and five distinct spectral states exist for black hole
X-ray binaries, depending on the classifying author. These states are
essentially accretion modes with distinct spectral and variability
characteristics.  The state of an object is, to first order,
correlated with its mass accretion rate, with often some hysteresis
(Miyamoto \& Kitamoto, 1995; Nowak, Wilms \& Dove 2002; Smith, Heindl \&
Swank 2002; Maccarone \& Coppi 2003).  For a more detailed review, we
refer to Nowak (1995) which concentrates more on the X-ray
spectroscopic features and van der Klis (1995) which concentrates more
on the variability features.

The three canonical spectral states are the low/hard state (typically
seen above below 2\% of the Eddington luminosity -- Maccarone 2003),
the high/soft state (typically seen between about 2 and 30\% of the
Eddington luminosity), and the very high state (typically seen above
about 30\% of the Eddington luminosity -- Nowak 1995).  The two
additional states are the quiescent state, which may just be the low
luminosity extension of the low/hard state, and the intermediate or
transition state, which shows spectral and timing properties
essentially the same as the very high state, but occurs at lower
luminosities, most notably as the source moves between the low/hard
state and the high/soft state (see Homan et al. 2001 for a discussion
of the intermediate state).  Active galactic nuclei seem to show
similar spectral states as black hole binaries at the same fractions
of the Eddington luminosity (Maccarone, Gallo \& Fender 2003), so the
extension of these results to intermediate mass black holes seems to
be a reasonable assumption.

\begin{figure}
\psfig{figure=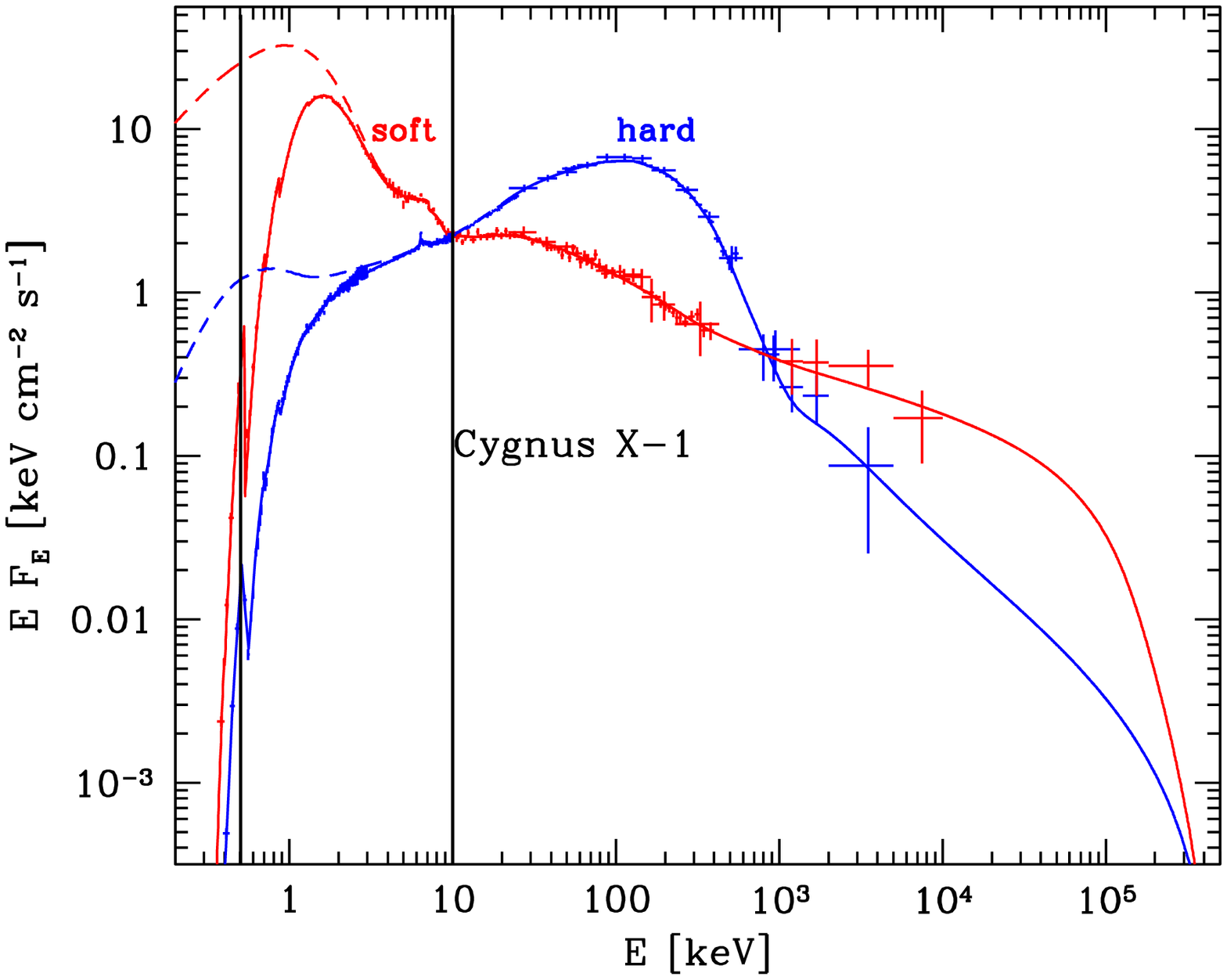,width=\columnwidth} 
\caption{ The spectral energy distributions of low/hard and high/soft
states of the canonical black hole X-ray binary, Cygnus X-1, adapted
from Figure 1 of Zdziarski \& Gierlinski (2004).  The data sets used
in compiling the figure are described in McConnell et al. (2000). The
y-axis shows the energy times the flux per unit energy ($F_E$), so
that the peak on the plot gives the energy where the power per decade
of the source peaks.  The solid lines indicates a model fit to the
measured flux, while the dashed curves indicates the model fit with
line-of-sight absorption removed.  The vertical lines indicate the
0.5-10 keV energy range typically used in {\em Chandra} observations.
The energy output in the high/soft state exceeds the low/hard state in
the {\em Chandra}-bandwidth ($\aplt 10$keV) and at energies above
$\sim 10^3$\,keV.
\label{fig:hardsoft}
}
\end{figure}

For illustrative purposes we present figure\,\ref{fig:hardsoft}, which
shows the X-ray spectrum of Cygnus X-1 in the soft state (top solid
curve for low $\aplt 10$keV and high energies $\apgt 10^3$keV) and in
the hard state (bottom curve in the {\em Chandra} band). The dashed
curves are corrected for absorption.

\subsubsection{The low/hard and quiescent states} 

The two lowest luminosity states are the quiescent state and the
low/hard state.  No clear indications exist that there are fundamental
differences in the accretion mode between these two states, and it has
been suggested that the quiescent state may be merely the lowest
luminosity low/hard state.  The X-ray spectra of low/hard states are
typically well fit by thermal Comptonization models with photon
indices, $\Gamma$ (defined such that the number of photons per unit
energy, $\frac{dN}{dE}=E^{-\Gamma})$ of 1.5--2.0 and roughly
exponential cutoffs at about 200 keV.  The quiescent state objects are
typically so faint that their spectral states cannot be well
constrained but they seem to have power law slopes consistent with
those of the low/hard state objects (Kong et al. 2002).  Additionally,
since their spectra have been measured only with soft X-ray imaging
telescopes, no spectral cutoffs have been seen from the quiescent
black holes.  Taking typical black hole low/hard state spectral
parameters (e.g. Nowak et al. 2002; McClintock et al. 2001) the
bolometric corrections from the {\em Chandra} band are factors of
$\sim$ 5--15.  If a X-ray source is in the low/hard state an observed
luminosity of $\apgt 10^{39}$\,erg\,s$^{-1}$ in the {\em Chandra}
bandpass indicates that the total luminosity should exceed $\sim
5\times10^{39}$\,erg\,s$^{-1}$; the black hole then should exceed
about 2000 \msun\, in order for this luminosity to be less than 2\% of
the Eddington luminosity.

Another key issue for converting mass accretion rate into observed
luminosity for the low luminosity states is the radiative efficiency
correction.  Geometrically thin accretion disks are generally believed
to be radiatively efficient.  Defining the radiative efficiency
$\eta=L/(\dot{M}c^2)$, it is generally found that $\eta=0.1$ for
non-rotating black holes' accretion disks, and a factor of $\sim4$
higher for disks around maximally rotating Kerr black holes.  On the
other hand, thick accretion disks, such as those thought to be
responsible for low/hard states, may be quite radiatively inefficient.
In the context of advection dominated accretion disks, it is found
that $\eta=10 ({\dot{M}}/\dot{M}_{\rm t})^2$ (Esin, McClintock \&
Narayan 1997), where $\dot{M}_{\rm t}$ is the mass accretion rate at
the luminosity of the state transition between the low/hard and
high/soft states.  The correlation between radio and X-ray luminosity
in low/hard state black hole binaries (Gallo, Fender \& Pooley 2003)
has been interpreted that the radio jet takes kinetic power from the
accretion flow in such a manner as to also produce a
$\eta\propto$$\dot{M}^2$ (Fender, Gallo \& Jonker 2003), although the
coefficient of proportionality was not well constrained by the data.

\subsubsection{The higher luminosity states} 

The other two (or three) states all have similar spectral properties.
They show dominant strong thermal emission at low energies, consistent
with a standard thermal accretion disk spectrum with a temperature of
$\sim1-2$ keV at the inner edge, and a weaker power law component with
a spectral index of $\Gamma\sim2.5$ and typically no observable
cutoff.  The very high state and the intermediate state typically show
a slightly higher fraction of their luminosity coming out in the power
law component, but in reality the difference between the two states is
best seen from the variability properties, as a strong $\sim0.5-10$ Hz
QPO is seen in the very high state, while the high/soft state
typically shows root-mean square fractional variability less than 5\%.
We take typical source parameters from Miller et al. (2001) who
presented tabulated fits to the RXTE spectra for an entire outburst of
the X-ray binary XTE J1748-288.  We then impose a low energy cutoff to
the power law at 2 keV, where the flux is dominated by the thermal
component, in order to keep the spectral model from diverging at low
energies (the exact location of this cutoff is unimportant so long as
it is in the photon energy range where the luminosity is dominated by
the thermal component).  We then compare the bolometric luminosity
from the spectral model to the bolometric luminosity in the {\it
Chandra} bandpass and find that about 75\% of the luminosity is within
the {\it Chandra} bandpass for the very high state and about 90\% for
the high/soft state.  Bolometric corrections are therefore not a major
source of uncertainty in converting between accretion rates and
luminosity for the high luminosity sources.

\subsection{Comments on past spectral state studies of ULXs}
Additionally, we wish to take the opportunity to make a few points
about the spectral states of X-ray binaries that may help explain
certain phenomenology of ULXs, and especially some points which seem
to contradict this picture, but in fact do not.  In particular, there
has been discussion of low/soft and high/hard spectral state
phenomenology.  We note first that with the {\em Chandra} bandpass, a
low/hard state will usually fit to a steeper power law spectral index
than will a high/soft state (e.g. Liu et al. 2002).  This is because
the peak of a 1-2 keV disk blackbody spectrum in power per decade will
be at about 3-6 keV.  The exponential cutoff of the spectrum will be
in the part of the {\em Chandra} bandpass where the effective area is
relatively low; with a few hundred total photons, as is typical for
ULX studies, the statistical weight of the spectral fit will be
dominated by the softer X-rays, where the spectrum is quite flat.
With very high signal-to-noise spectra, where one can actually rule
out a power law model for the high/soft states or a disk blackbody
model for the low/hard states; only then is the spectral state
identification useful for comparison with the black hole binaries.  We
note that Kubota, Done \& Makishima (2002) have also pointed out that
ULX spectra should not be compared with the low/hard states of
accreting black holes, but rather with the more high luminosity
states.

A second key related point is that in the high/soft state, the
accretion disk seems to extend in to a constant inner radius (see
e.g. Sobczak et al. 2000), which is consistent with being the
marginally stable orbit around the black hole.  The luminosity
variations are then determined almost entirely by changes in the disk
temperature.  In the very high state, the disk may change size a bit
more, but still the luminosity rises are correlated primarily with
increases in the temperature of the disk and increases in the ratio of
the power law component to the blackbody component.  That is to say,
even in normal X-ray binaries, at high X-ray luminosities, the spectra
frequently harden as the luminosity rises, but without the qualitative
change in behavior associated with state transitions.

Some recent work on bright well-studied ULXs show slight deviations
from the simplistic picture presented for ULXs.  In particular, the
hardness ratios of a bright source in the Antennae galaxies are
sometimes positively correlated with flux and sometimes inversely
correlated with flux, with some indications that loops are traced out
in a hardness versus intensity diagram (Zezas et al. 2003).  This is
consistent with behavior seen in the Galactic microquasar GRS 1915+105
(Fender \& Belloni 2004), where within the very high state, such loops
are quite often seen.  This may be revealing some quite interesting
physics, as discussed in the reference above, but is unlikely to have
a major affect on the bolometric corrections we have described here.

\subsection{Duty cycles, recurrence timescales, and predicted appearance} 

We note further that in the outbursts of bright X-ray transients, the
bulk of the mass accretion will occur during the high luminosity
phases of the outburst.  Computations of the outburst duty cycle, then
can assume radiatively efficient accretion, and can assume the
exponential decay continues down to arbitrarily small luminosity (in
reality, soft X-ray transients seem not to drop below about $10^{-9}$
of the Eddington limit -- Garcia et al. 2001) without seriously
affecting the parameter values.



\newpage

\section*{Erratum}

In the paper ``Intermediate Mass Black Holes in Accreting Binaries:
Formation, Evolution and Observational Appearance'' by Simon F.\,
Portegies Zwart, Jasinta Dewi and Tom Maccarone (MNRAS October [2004])
contains an error causing the frequency of the gravitational wave
sources to be too large by a factor $\pi$.  This affects the
discussion in \S\,4.1 and the numbers in the right most column of
Tab.\, 1. If instead of a distance of 10\,kpc one assumes a distance
of 1\,kpc the new numbers in the right most column of Tab.\, 1 become
1200\,Myr, 0, 10, 0, 0, 0, 0, 0, 2, 0, 0, 0, 0, 1\,Myr, from the top
row to the bottom row.

For clarity we present a new figure 4 based on a distance of 1\,kpc.

\begin{figure}
\psfig{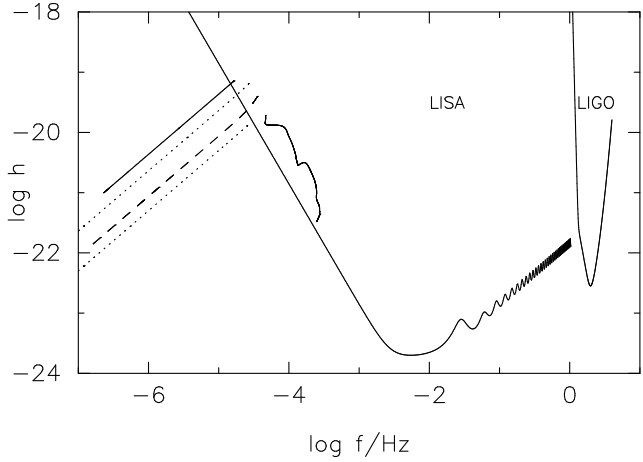} 
\caption{Updated figure 4 assuming a distance of 1\,kpc. Further
         caption is as in the original published version.  }
\end{figure}

We thank Gijs Nelemans for pointing out the ommission.

\bsp
\label{lastpage}

\end{document}